\def \JMP {{ J. Math. Phys. }} 
\def \NP {{ Nucl. Phys.} }

\def \bc {\begin{center}}
\def \ec {\end{center}}

\def \bfr {\begin{flushright}}
\def \efr {\end{flushright}}

\def \ba{\begin{array}}
\def \ea{\end{array}}

\def \bea {\begin{eqnarray}}
\def \eea {\end{eqnarray}}

\def \be {\begin{equation}}
\def \ee {\end{equation}}

\def\ni{\noindent}
\def\nn{\nonumber}

\def\f{\frac}
\def\l[{\left[}
\def\r]{\right]}

\def\cL{{\cal L}}

\def\kbar{{\mathchar'26\mkern-9mu\lambda}}

\documentstyle[11pt,a4]{article}

\begin{document}

\centerline{}\vskip 1.5cm
\begin{center} 
{\large {\bf Promoting finite to infinite symmetries: 
the $3+1$-dimensional analogue of the Virasoro algebra and higher-spin fields}}
\end{center}
\bigskip
\centerline{ {\sc M. Calixto}\footnote{E-mail: pymc@swansea.ac.uk 
/ calixto@ugr.es}  }
\bigskip

\begin{center}
{\it Department of Physics, University of Wales Swansea, Singleton Park, 
Swansea, SA2 8PP, U.K.}\\ and \\
{\it Instituto Carlos I de F\'\i sica Te\'orica y Computacional, Facultad
de Ciencias, Universidad de Granada, Campus de Fuentenueva, 
Granada 18002, Spain.} 
\end{center}

\bigskip

\bigskip
\begin{center}
{\bf Abstract}
\end{center}
\small          

\begin{list}{}{\setlength{\leftmargin}{3pc}\setlength{\rightmargin}{3pc}}
\item 
Infinite enlargements of finite pseudo-unitary symmetries are  
explicitly provided in this letter. The particular case of 
$u(2,2)\simeq so(4,2)\oplus u(1)$ constitutes a 
(Virasoro-like) infinite-dimensional 
generalization of the $3+1$-dimensional 
conformal symmetry, in addition to matter fields with 
all conformal spins. These algebras provide  
a new arena for integrable field models in higher dimensions; for example, 
Anti-de Sitter and conformal 
gauge theories of higher-$so(4,2)$-spin fields. 
A proposal for a non-commutative geometrical  
interpretation of space is also outlined.

\end{list}

\normalsize

\vfill 
\ni PACS: 02.20.Tw, 11.25.Hf, 04.50.+h 

\ni KEYWORDS: infinite symmetry, conformal field theory, induced gravity, 
integrability, noncommutative space.


\newpage

\noindent Conformal symmetry $SO(D,2)$ in $D$ 
space-time dimensions is said to be 
finite-dimensional except for $D=2$, in which case it can be enlarged to 
two copies (the right- and 
left- moving modes) of the infinite-dimensional Witt algebra (center-less 
Virasoro algebra):
\be
\left[L_m,L_n\right]=(m-n)L_{m+n}\,,\;\;\;m,n\in Z.\label{Vir}
\ee
The infinite and non-Abelian character of this algebra makes possible 
the exact solvability of conformally-invariant (quantum and statistical) 
non-linear field theories in one and two dimensions, and helps 
with systems in higher 
dimensions which, in some essential respects, are one- or two-dimensional 
(e.g. String Theory). There can be no doubt that a higher-dimensional 
counterpart of this infinite symmetry would provide novel insights into 
the analysis and formulation of new non-linear integrable conformal 
field theories and, more ambitiously, a potential 
starting point to formulate quantum gravity models in realistic dimensions. 
We shall show how such an analogue 
can be found inside a more general family of infinite-dimensional 
algebras based on pseudo-unitary symmetries.

The Witt algebra (\ref{Vir}) can be also seen as a sort of ``analytic 
continuation'' (see e.g. \cite{Fradkin2,Pope} for similar concepts) of 
$su(1,1)=\{L_n,|n|\leq 1\}$, that is, an extension beyond the 
``wedge'' $|n|\leq 1$ by revoking this restriction to $n\in Z$. 
Analytic continuations of pseudo-unitary algebras $u(N_+,N_-)$ can be 
given as well and they provide, from 
this point of view, infinite-dimensional generalizations of 
the conformal symmetry in realistic dimensions for the particular case 
of $u(2,2)\simeq so(4,2)\oplus u(1)$. 

Indeed, let us denote by 
${X}_{\alpha\beta},\, \alpha,\beta=1,\dots,N\equiv N_++N_-$, 
the $U(N_+,N_-)$ Lie-algebra of (step) generators with commutation relations:
\be
\left[{X}_{\alpha_1\beta_1},{X}_{\alpha_2\beta_2}\right]=
(\eta_{\alpha_1\beta_2}{X}_{\alpha_2\beta_1}-
\eta_{\alpha_2\beta_1}{X}_{\alpha_1\beta_2})\,,\label{pun}
\ee 
where $\eta={\rm diag}(1,\stackrel{N_+}{\dots},1,-1,
\stackrel{N_-}{\dots},-1)$ is used to raise and lower indices (see a 
clarifying Note at the end of the paper). 
For example, 
for $u(1,1)$ we have: $L_1=X_{12}, L_{-1}=X_{21}, 
L_0=\frac{1}{2}(X_{22}+X_{11})$. It is of importance that the commutation 
relations  (\ref{pun}) can also be written as
\be
\left[L^I_m, L^J_n\right]=\eta^{\alpha\beta}(J_\alpha m_\beta-
I_\alpha n_\beta)L^{I+J-\delta_\alpha}_{m+n}\,,\label{inf}
\ee
where the the lower index 
$m$ of $L$ now symbolizes a integral upper-triangular $N\times N$ matrix and 
the upper (generalized spin) index $I\equiv(I_1,\dots,I_N)$  represents 
a $N$-dimensional vector which, in general, can be taken to 
lie on a half-integral lattice; we denote by $m_\alpha\equiv
\sum_{\beta>\alpha} m_{\alpha\beta}-\sum_{\beta<\alpha} m_{\beta\alpha}$ and 
$\delta_\alpha\equiv (\delta_{\alpha,1},\dots,\delta_{\alpha,N})$. Note that, 
the generators  ${L}^I_m$ are labeled by $N+N(N-1)/2=N(N+1)/2$ indices, 
in the same way as wave functions $\psi^I_m$ in the carrier space 
of irreps of $U(N)$. There are many possible ways of embedding the 
$u(N_+,N_-)$ generators (\ref{pun}) inside (\ref{inf}), as there are also 
many possible choices of $su(1,1)$ inside (\ref{Vir}). However, 
a ``canonical'' choice is 
\be
X_{\alpha\beta}\equiv 
L^{\delta_\alpha}_{x_{\alpha\beta}}\,, \;\;\;
x_{\alpha\beta}\equiv {\rm sign}(\beta-\alpha)
\sum_{\sigma={\small{\rm Min}}(\alpha,\beta)}^{{\small{\rm 
Max}}(\alpha,\beta)-1} x_{\sigma,\sigma+1}\,,\label{embedding}
\ee
where $x_{\sigma,\sigma+1}$ 
denotes an upper-triangular matrix with zero 
entries, except for one at the $(\sigma,\sigma+1)$-position, that is 
$(x_{\sigma,\sigma+1})_{\mu\nu}=\delta_{\sigma,\mu}\delta_{\sigma+1,\nu}$ 
(we set $x_{\alpha\alpha}\equiv 0$). 
For example, the $u(1,1)$ Lie-algebra generators correspond to:
\be
X_{12}=L^{(1,0)}_{\left(\begin{array}{cc} 0 &1\\ 0&0\ea\right)},\;\;
X_{21}=L^{(0,1)}_{\left(\begin{array}{cc} 0 &-1\\ 0&0\ea\right)},\;\;
X_{11}=L^{(1,0)}_{\left(\begin{array}{cc} 0 &0\\ 0&0\ea\right)},\;\;
X_{22}=L^{(0,1)}_{\left(\begin{array}{cc} 0 &0\\ 0&0\ea\right)}.
\ee

Relaxing the restrictions $m=x_{\alpha\beta}$ in (\ref{embedding}) 
to arbitrary integral upper-triangular matrices $m$ leads to the  
following infinite-dimensional algebra (as can be seen from (\ref{inf}))
\be
\l[L^{\delta_\alpha}_{m}, 
L^{\delta_\beta}_{n}\r]= m^\beta L^{\delta_\alpha}_{m+n}-
n^\alpha L^{\delta_\beta}_{m+n}\,,\label{difeounm}
\ee
which we shall denote by $\cL_{\infty}^{(1)}(u(N_+,N_-))$. 
It is easy to see that, for $u(1,1)$, this ``analytic continuation'' 
leads to two Virasoro sectors: $L_{m_{12}}\equiv L^{(1,0)}_m,\,
\bar{L}_{m_{12}}\equiv L^{(0,1)}_m$. Its $3+1$ dimensional counterpart 
$\cL_{\infty}^{(1)}(u(2,2))$ contains four non-commuting Virasoro-like sectors 
$\cL_{\infty}^{(1_\alpha)}(u(2,2))=\{L^{\delta_\alpha}_{m}\}
,\,\alpha=1,\dots,4$ which, in their turn, 
hold three genuine Virasoro sectors for $m=k u_{\alpha\beta},\, 
k\in Z,\, \alpha<\beta=2,\dots,4$, where $u_{\alpha\beta}$ denotes an 
upper-triangular matrix with components 
$(u_{\alpha\beta})_{\mu\nu}=\delta_{\alpha,\mu}\delta_{\beta,\nu}$. 
In general, $\cL_{\infty}^{(1)}(u(N_+,N_-))$ 
contains $N(N-1)$ distinct and non-commuting 
Virasoro sectors, and holds $u(N_+,N_-)$ as the {\it maximal 
finite-dimensional subalgebra}.

The algebra $\cL_{\infty}^{(1)}(u(N_+,N_-))$ can be seen as the {\it minimal} 
infinite continuation of $u(N_+,N_-)$ representing the diffeomorphism 
algebra  diff$(N)$ of the corresponding $N$-dimensional manifold (locally the 
Minkowski space-time for $u(2,2)$). Indeed, the algebra (\ref{difeounm}) 
formally coincides with the algebra of vector fields 
$L^\mu_{f(y)}=f(y)\frac{\partial}{\partial y_\mu}$, where 
$y=(y_1,\dots,y_N)$ denotes a local system of coordinates and $f(y)$ 
can be expanded in a plane wave basis, such that 
$L^\mu_{\vec{m}}=e^{im^\alpha y_\alpha}
\frac{\partial}{\partial y_\mu}$ 
constitutes a basis of vector fields for  the so called 
generalized Witt algebra 
\cite{Ree}, of which there are studies about its representations (see e.g. 
\cite{Rao,Fabbri,Larsson99}). Note that, for us, the $N$-dimensional 
lattice vector $\vec{m}=(m_1,\dots,m_N)$ is constrained by 
$\sum_{\alpha=1}^N m_\alpha=0$ (see the definition of $m_\alpha$ in 
paragraph after Eq. \ref{inf}), which 
introduces some novelties as regards the Witt algebra. In fact, the algebra 
(\ref{difeounm}) can be split into one ``temporal'' piece, constituted by 
an Abelian ideal generated by $\check{L}^N_m\equiv \eta_{\alpha\alpha} 
L^{\delta_\alpha}_{m}$, and a ``residual'' symmetry generated by the 
spatial diffeomorphisms 
\be
\check{L}^j_m\equiv\eta_{jj} 
L^{\delta_j}_{m}-\eta_{j+1,j+1} L^{\delta_{j+1}}_{m},\,j=1,\dots,N-1\,\, 
({\rm no \ sum \ on \ } j)\,,
\ee
which act semi-directly on the temporal part. More precisely, the 
commutation relations (\ref{difeounm}) in this new basis adopt the following 
form:
\bea
\l[ \check{L}^j_m,\check{L}^k_n\r] &=& \check{m}^k \check{L}^j_{m+n} -
\check{n}^j \check{L}^k_{m+n}\,,\nn\\
\l[ \check{L}^j_m,\check{L}^N_n\r] &=&  -\check{n}^j \check{L}^N_{m+n}\,,
\label{inftempesp}\\
\l[ \check{L}^N_m,\check{L}^N_n\r] &=& 0\,,\nn
\eea
where $\check{m}_k\equiv m_k-m_{k+1}$. Only for $N=2$, the last 
commutator admits a central extension of the form  
$\sim n_{12}\delta_{m+n,0}$ 
compatible with the rest of commutation relations 
(\ref{inftempesp}). This result amounts to the fact that the 
(unconstrained) diffeomorphism algebra diff$(N)$ does not admit any 
non-trivial central extension except when $N=1$ \cite{nocentral}.

Additionally, after the restriction 
$I=\delta_\alpha$ in (\ref{embedding}) is also relaxed to arbitrary 
half-integral lattice vectors $I$, the commutation 
relations (\ref{inf}) define a {\it 
higher-$u(N_+,N_-)$-spin algebra} $\cL_{\infty}(u(N_+,N_-))$ (in a sense 
similar to  that of Ref. \cite{Fradkin}), which 
contains $\cL_{\infty}^{(1)}(u(N_+,N_-))$ as a subalgebra as well as 
all {\it matter fields} $L^I_m$ with all $u(N_+,N_-)$-spins $I$. 

The quantization procedure for the algebra (\ref{inf}) entails unavoidable 
renormalizations (mainly due to ordering problems) and central extensions 
like:
\be
\left[\hat{L}^I_m, \hat{L}^J_n\right]=
\hbar\eta^{\alpha\beta}(J_\alpha m_\beta-
I_\alpha n_\beta)\hat{L}^{I+J-\delta_\alpha}_{m+n}+ O(\hbar^3) + 
\hbar^{(\sum_{\alpha=1}^N{I_\alpha+J_\alpha})}c^{(I,J)}(m)\delta_{m+n,0}\hat{1}
\,,\label{infq}
\ee
where $\hat{1}\sim \hat{L}^0_0$ denotes a central generator and 
$c^{(I,J)}(m)$ are central charges. The higher order terms $O(\hbar^3)$ 
can be captured in a classical construction by extending the 
classical (Poisson-Lie) bracket (\ref{inf}) to the Moyal bracket (see 
\cite{infdimal} for more information on Moyal deformation). 

Central extensions provide the essential ingredient  
required to construct invariant geometric 
action functionals on coadjoint orbits of the corresponding groups. 
When applied to the infinite continuation 
(\ref{infq}) of $u(2,2)$, this would lead  
to Wess-Zumino-Witten-like models for {\it induced gravities in $3+1$ 
dimensions}, as  happens for the Virasoro and ${\cal W}$ algebras 
in $1+1D$ (see e.g. \cite{Nissimov}). 
The minimal coupling to matter could be done just 
by adding the {\it full} set of conformal 
fields $\hat{L}^I_m$ with all $u(2,2)$-spins $I=(I_1,I_2,I_3,I_4)$.

The higher-$u(N_+,N_-)$-spin algebras (\ref{infq})  provide the arena 
for new non-linear integrable quantum field models in higher dimensions, the 
particular cases $u(2,2)\simeq so(4,2)\oplus u(1)$ and 
$u(4)\simeq so(6)\oplus u(1)$ being also a potential guiding principle 
towards the still unknown ``M-theory''.

An additional point which is also worth-mentioning, is that 
the algebra (\ref{infq}) is actually a member 
of a $N$-parameter family $\tilde{\cL}_{\vec{\rho}}(u(N_+,N_-)),\,
\vec{\rho}\equiv (\rho_1,\dots,\rho_N)$ of non-isomorphic algebras of 
$U(N_+,N_-)$ tensor operators (see \cite{infdimal}), 
the classical limit $\hbar\to 0, \rho_\alpha\to \infty$  corresponding to 
the classical (Poisson-Lie) algebra $\cL_{\infty}(u(N_+,N_-))$ with 
commutation relations (\ref{inf}). A very interesting feature of 
$\tilde{\cL}_{\vec{\rho}}(u(N_+,N_-))$ is that it {\it collapses} to 
${\rm Mat}_{d}({C})$ (the full matrix algebra of $d\times d$ 
complex matrices) whenever the (complex) parameters 
$\rho_\alpha$ coincide with the eigenvalues $q_\alpha$ 
of the Casimir operators $C_\alpha$ of $u(N_+,N_-)$ in a 
$d$-dimensional irrep $D_{\vec{q}}$ of 
$u(N_+,N_-)$. This fact can provide finite ($d$-points) 
`fuzzy' or `cellular' descriptions of the non-commutative counterpart 
of AdS$_5$ (a desirable property as regards finite models of quantum gravity) 
when applying the ideas of {\it non-commutative geometry} (see e.g. 
\cite{Connes}) to $\tilde{\cL}_{\vec{\rho}}(u(2,2))$. 

Let us illustrate 
briefly this phenomenon with the simple example of the algebra 
$C^\infty(T^2)$ of smooth functions $L_{\vec{m}}=
e^{\frac{2\pi i}{\ell} \vec{m}\cdot\vec{y}}$ on a torus $T^2$, where 
$\vec{y}=(y_1,y_2)$ denote the coordinates (modulo $\ell$) and  
$\vec{m}=(m_1,m_2)$ are a pair of integers. The ordinary product of functions 
$L_{\vec{m}}\cdot L_{\vec{n}}= L_{\vec{m}+\vec{n}}$ 
defines $C^\infty(T^2)$ as an associative 
and commutative algebra. Yet, it must be emphasized that 
this product is actually a limiting case of a 
more fundamental (quantum) associative and {\it non-commutative} 
$\star$-product $\hat{L}_{\vec{m}}\star \hat{L}_{\vec{n}}=
e^{2\pi i\frac{\kbar^2}{\ell^2}\vec{m}\times\vec{n}} 
\hat{L}_{\vec{m}+\vec{n}}$, where $\vec{m}\times\vec{n}\equiv m_1n_2-m_2n_1$, 
$\kbar$ is a parameter with dimensions 
of length (e.g. the Planck length $\sqrt{G\hbar}$) and $\hat{L}_{\vec{m}}$ 
denotes a {\it symbol} (the non-commutative counterpart 
of the function $L_{\vec{m}}$). The commutator of 
two symbols is defined as:
\be
\left[\hat{L}_{\vec{m}},\hat{L}_{\vec{n}}\right]\equiv
\hat{L}_{\vec{m}}\star\hat{L}_{\vec{n}}-\hat{L}_{\vec{n}}
\star\hat{L}_{\vec{m}}=
2i\sin\left(2\pi \frac{\kbar^2}{\ell^2} \vec{m}\times\vec{n}\right)
\hat{L}_{\vec{m}+\vec{n}} \,.\label{Moyal}
\ee
Note that, when the surface of the torus $\ell^2$ contains an integer 
number $q$ of times the {\it minimal cell} $\kbar^2$ (that is, 
$\ell^2=q \kbar^2$), the infinite-dimensional algebra 
(\ref{Moyal}) collapses to a finite-dimensional matrix algebra: the Lie 
algebra of the unitary group $U(q/2)$ for $q$ even or 
$SU(q)\times U(1)$ for $q$ odd (see \cite{Fairlie}). 
In fact, taking the quotient in (\ref{Moyal}) by the equivalence 
relation  $\hat{L}_{\vec{m}+q\vec{a}}\sim 
\hat{L}_{\vec{m}},\,\forall \vec{a}\in Z\times Z$, it can be seen that 
the following identification $\hat{L}_{\vec{m}}=\sum_{k}
e^{\frac{2\pi i}{q}m_1k}X_{k,k+m_2}$ 
implies a change of basis in the step-operator algebra (\ref{pun}) 
of $U(q)$. 

Thinking of $\rho=\frac{\ell^2}{\kbar^2}$ as a `density of points', 
we can conclude that: for the critical values $\rho_c=q\in Z$, the Lie algebra 
(\ref{Moyal}) ---which we denote by $C^\star_\rho(T^2)$--- is {\it finite}; 
that is, the quantum analogue of the torus has a `finite number $q$ of 
points'. It is in this sense that we talk about a 
`cellular structure of the space'. Moreover, given the basic commutator 
$\left[ y_1,y_2\right]=-i\kbar^2/\pi$, this cellular structure is a 
consequence of the absence of localization expressed by the Heisenberg 
uncertainty relation $\Delta y_1 \Delta y_2 \geq \kbar^2/(2\pi)$. 

In the (classical) limit of large number of points $\rho\to\infty$ and 
$\kbar\to 0$ (such that $\ell^2=\rho\kbar^2$ remains finite) we 
recover the original (commutative) geometry on the torus. 
For example, it is easy to see that 
\be
\lim_{{\small \ba{c}\rho\to \infty\\ \kbar\to 0\ea}} 
\frac{i\pi}{\kbar^2}\left[\hat{L}_{\vec{m}},\hat{L}_{\vec{n}}\right]
=\frac{4\pi^2}{\ell^2}\vec{n}\times\vec{m}\,\hat{L}_{\vec{m}+\vec{n}}
\ee
coincides with the (classical) Poisson bracket 
$\left\{ L_{\vec{m}},L_{\vec{n}}\right\}=
\Upsilon_{jk}\frac{\partial L_{\vec{m}}}{\partial y_j}
\frac{\partial L_{\vec{n}}}{\partial y_k}$ of functions in 
$C^\infty(T^2)$,  where 
$\Upsilon_{2\times 2}\equiv
\left(\begin{array}{cc} 0 & 1 \\ 
-1 &0\ea\right)$ denotes the symplectic form on the torus. 

\vskip 1cm
I thank the University of Granada for a Post-doctoral grant and the 
Department of Physics of Swansea for its hospitality. I wish to acknowledge 
helpful comments of T.A. Larsson who brought the papers 
\cite{Ree,Rao,Fabbri,Larsson99}, among others, to my attention. 

\vskip 1cm
{\bf Note added:} when working in the 
complexification of $u(N_+,N_-)$, the Lie 
algebras  $u(N_+,N_-)=\{X_{\alpha\beta}\}$ and $u(N)=
\{X_{\alpha}^\gamma\}$ are related by ${X}_{\alpha\beta}=
\eta_{\beta\gamma}{X}_{\alpha}^\gamma$; however, this relation 
does not mean that $u(N_+,N_-)$ and $u(N)$ are isomorphic as such. Indeed, 
let us take the simple case of
\be
G(\kappa)=\left\{ g= \left( \begin{array}{cc} z_1&\kappa\bar{z}_2 \\ z_2 &
\bar{z}_1\end{array}
\right) ,z_i,\bar{z}_i \in C/ \det(g)=|z_1|^2-\kappa |z_2|^2=1 \right\}\,,
\ee
which reduces to $G(1)=SU(1,1)$ and $G(-1)=SU(2)$ for $\kappa=\pm 1$. 
Let us choose the 
following system of complex coordinates: $z\equiv\f{z_2}{z_1}, \,
\bar{z}\equiv\f{\bar{z}_2}{\bar{z}_1}, \,\xi\equiv\f{z_1}{|z_1|}$; 
they correspond to the standard stereographic projection coordinates 
$z,\bar{z}$ of the sphere (resp. hyperboloid) on the complex plane $C$ 
(resp. unit disk) for $SU(2)$ (resp. $SU(1,1)$), and $\xi$ is 
the Cartan phase.   
After a little bit of algebra, one can 
see that there is a change of basis between both ($su(2)$ and $su(1,1)$) 
{\it abstract} Lie algebras of left-invariant vector fields $X$ of the form: 
$X_z^{SU(2)}=X_z^{SU(1,1)},\,X_{\bar{z}}^{SU(2)}=-X_{\bar{z}}^{SU(1,1)},
\,X_\xi^{SU(2)}=X_\xi^{SU(1,1)}$ (see \cite{symplin} for more 
details); however, as stated above, it does not 
mean that $su(2)$ and $su(1,1)$ are isomorphic!. As abstract algebras, 
the main difference arises from the point of view of representations, when 
$(X_z^{SU(1,1)})^\dag=X_{\bar{z}}^{SU(1,1)}$ whereas 
$(X_z^{SU(2)})^\dag = -X_{\bar{z}}^{SU(2)}$.


\begin{thebibliography}{99}
\bibitem{Fradkin2} E.S. Fradkin and V.Y. Linetsky, \JMP {\bf 32}, 1218 (1991).
\bibitem{Pope} C.N. Pope, X. Shen and  L.J. Romans, \NP {\bf B339}, 191 (1990).
\bibitem{Ree} R. Ree, Trans. Amer. Math. Soc. {\bf 83}, 510 (1956).
\bibitem{Rao} S. Eswara Rao and R.V. Moody, Commun. Math. Phys. {\bf 159}, 
239 (1994).
\bibitem{Fabbri} M. Fabbri and R.V. Moody, Commun. Math. Phys. {\bf 159}, 
1 (1994).
\bibitem{Larsson99} T.A. Larsson, Commun. Math. Phys. {\bf 201}, 
461 (1999).
\bibitem{nocentral} E. Ramos, C.H. Sah and R.E. Shrock, 
J. Math. Phys. {\bf 31}, 1805 (1990).
\bibitem{Fradkin} E.S. Fradkin and M.A. Vasiliev, 
Ann. Phys. (NY) {\bf 77}, 63 (1987).
\bibitem{infdimal} M. Calixto, J. Phys. {\bf A33} (Math. Gen.) L69 (2000).
\bibitem{Nissimov} E. Nissimov, S. Pacheva and I. Vaysburd, 
Phys. Lett. {\bf B288}, 254 (1992). 
\bibitem{Connes} A. Connes, Noncommutative Geometry, Academic Press (1994).
\bibitem{Fairlie} D.B. Fairlie and P. Fletcher, J. Math. Phys. {\bf 31}, 1088 
(1990).
\bibitem{symplin} M. Calixto, V. Aldaya and J. Guerrero, Int. J. Mod. Phys. 
{\bf A13}, 4889 (1998).
\end{thebibliography}
\end{document}